\documentclass[a4paper,11pt]{article}
\pdfoutput=1
\usepackage{jheppub}
\usepackage{amsmath,amsfonts}
\usepackage{amssymb}
\usepackage{booktabs}
\usepackage{comment}
\usepackage{bbold}

\usepackage{multirow}

\usepackage{float}
\usepackage{subfig}
\usepackage{placeins}

\newcommand{\Q}{\mathcal Q}

\newcommand{\A}{\mathcal A}
\newcommand{\B}{\mathcal B}
\newcommand{\C}{\mathcal C}
\newcommand{\M}{\mathcal M}
\newcommand{\E}{\mathcal E}
\newcommand{\Z}{\mathcal Z}
\newcommand{\Y}{\mathcal Y}
\newcommand{\D}{\mathcal D}
\newcommand{\pv}{$p$\,-value}
\newcommand{\pvs}{$p$\,-values}

\graphicspath{ {figs/} }

\definecolor{lightgray}{rgb}{0.7,0.7,0.7}

\newcommand{\be}{\begin{equation}}
\newcommand{\ee}{\end{equation}}
\newcommand{\bea}{\begin{eqnarray}}
\newcommand{\eea}{\end{eqnarray}}

\newcommand{\eps}{\epsilon}

\title{Realistic GUT Yukawa Couplings from a Random Clockwork Model.}
\author[a]{Gero von Gersdorff}
\affiliation[a]{Department of Physics, Pontif\'icia Universidade Cat\'olica do Rio de Janeiro, Rua Marqu\^es de S\~ao Vicente 225, Rio de Janeiro, Brazil}

\emailAdd{gersdorff@puc-rio.br}

\abstract{We present realistic models of flavor in $SU(5)$ and $SO(10)$ grand unified theories (GUTs). The models are renormalizable and do not require any exotic representations in order to accommodate the necessary GUT breaking effects in the Yukawa couplings. They are based on a simple clockwork Lagrangian whose structure is enforced with just two (one) vectorlike $U(1)$ symmetries in the case of $SU(5)$ and $SO(10)$ respectively. The inter-generational hierarchies arise spontaneously from products of matrices with order one random entries.}

\begin{document}
\maketitle

\section{Introduction}

The puzzling flavor structure of the Standard Model (SM)  has inspired a lot of model building over the past decades. The  large ratios of masses and mixing parameters displayed in table \ref{tab:exp}, commonly referred to as flavor hierarchies, cannot be the result of a generic UV theory with order-one couplings.

Another feature of the SM matter sector is the peculiar structure of its five gauge  representations, which strongly hints at some unified gauge group with only
 two $SU(5)$ or one $SO(10)$ representation.
 These so-called grand-unified theories (GUTs) in turn strongly motivate the extension of the SM to its minimally supersymmetric version, the MSSM, as the latter provides an accurate unification of the gauge couplings near a scale of the order of $10^{16}$ GeV, commonly referred to as the GUT scale.
 However, the unification of couplings is much less impressive in the Yukawa sector. For instance, in $SU(5)$ unification, one finds the GUT-scale relation 
 $\Y_d=\Y_e^T$ for the Yukawa couplings matrices of the charged-lepton and down-quark sectors.
 A quick glance at table \ref{tab:exp} (which shows the couplings at the GUT scale) reveals that this is certainly not the case. Even though  supersymmetric threshold corrections are more important than in the gauge sector (and more  dependent on the superpartner spectrum), it is very hard to attribute the large differences in the couplings to this alone. 
 Typically it is possible to adjust the spectrum such that only some but not all of the Yukawa couplings become unified.
 Certain GUT breaking effects at the high scale thus seem to be unavoidable.
 Some efforts have been made  to include such breaking effects, for instance via exotic Higgs representations \cite{Georgi:1979df},~nonrenormalizable operators \cite{Ellis:1979fg,Antusch:2014poa},  vectorlike representations \cite{Murayama:1991ew}, or combinations of these \cite{Altarelli:2000fu}.
 In minimal $SO(10)$ unification, all Yukawa couplings are predicted to be exactly equal, $\Y_u=\Y_d=\Y_e^T$. The necessary GUT breaking effects now become even larger, as a comparison between the different columns of table \ref{tab:exp} shows.

\begin{table}[htb]
\centering
\begin{tabular}{cc|  cc | cc |c c| c cc}
\toprule
$y'_u$	&	$3\times 10^{-6}$	&	$y''_d$	&	 $0.5 \times 10^{-5}$		&	$\theta_{13}$	&	$3.3\times 10^{-3}$	&	$y''_e$	&	$2.0 \times 10^{-6}$	&	$\sin^2\theta_{13}$	&	0.02	\\
$y'_c$	&	$1.4\times 10^{-3}$	&	$y''_s$	&	 $1.0 \times 10^{-4}$ 	&	$\theta_{23}$	&	$3.7\times 10^{-2}$	&	$y''_\mu$	&	$4.5 \times 10^{-4}$	&	$\sin^2\theta_{23}$	&	0.5	\\
$y'_t$	& 	$0.55$			&	$y''_b$	&	 $0.6 \times 10^{-2}$		&	$\theta_{12}$	&	$0.23$			&	$y''_\tau$	&	$1.0 \times 10^{-2}$	&	$\sin^2\theta_{12}$	&	0.3\\
\bottomrule
\end{tabular}
\caption{Quark and Lepton data at the GUT scale in the MSSM \cite{Antusch:2013jca}.  
Here, $y'_i=y_i\sin\beta$ and $y''_i=y_i\cos\beta$.
The values are representative, as they depend  on the supersymmetric threshold corrections, and indirectly on $\tan\beta$ via the renormalization group running.
}\label{tab:exp}. 
\end{table}
 
The Clockwork (CW) mechanism, originally formulated in \cite{Choi:2015fiu,Kaplan:2015fuy} in the context of the Relaxion, 
has soon been recognized as a general framework for constructing natural hierarchies  \cite{Giudice:2016yja}. 
In the flavor sector, it has been applied to explain the smallness of neutrino masses 
\cite{Ibarra:2017tju,Banerjee:2018grm,Hong:2019bki,Kitabayashi:2019qvi,Hambye:2016qkf} as well as the charged flavor hierarchies mentioned above 
\cite{Burdman:2012sb,vonGersdorff:2017iym,Patel:2017pct,Alonso:2018bcg,Smolkovic:2019jow,deSouza:2019wji}.
The CW Lagrangian is basically a one dimensional lattice ("theory space") of nearest-neighbor interactions enforced by some symmetry.
The generation of hierarchies  can be attributed to a controlled localization of zero modes  towards the boundaries of theory space.
However, it has also been pointed out \cite{Craig:2017ppp,deSouza:2019wji,Tropper:2020yew} that when the parameters in the CW Lagrangian are chosen at random, sharp localization of the zero modes occurs in the bulk of theory space, still leading to hierarchical suppression of couplings. 
Moreover, when the couplings of the CW model become  $3\times 3$ matrices in flavor space, the three zero modes spontaneously localize at different points in the lattice, and generate inter-generational hierarchies (the vertical hierarchies in table \ref{tab:exp}) \cite{deSouza:2019wji}.
On a technical level, this is closely related to a peculiar property of products of $N$ random matrices, which feature a very hierarchical spectrum despite their matrix elements being all of order one \cite{vonGersdorff:2017iym}.
 
In this paper, we are going to build models of natural flavor hierarchies along the lines of Refs.~\cite{vonGersdorff:2017iym,deSouza:2019wji} in supersymmetric $SU(5)$ and $SO(10)$ unification. The necessary GUT breaking effects are naturally present in these models, as the symmetries of the CW Lagrangian allow for renormalizable Yukawa couplings of the vectorlike fields (CW gears) with the GUT-breaking Higgs field(s). The flavor hierarchies and Yukawa unification thus have a common origin in the framework of a completely renormalizable model without any exotic Higgs or matter representations.

This paper is organized as follows. In section \ref{sec:models} we describe our models and briefly review the mechanism of flavor hierarchies. In section \ref{sec:sims} we perform comprehensive scans over the parameter space of these models, and quantify how well the SM flavor structure is reproduced. 
Some phenomenological considerations are given in section \ref{sec:pheno}, and in section \ref{sec:conclusions} we present our conclusions.

\section{The Model}
\label{sec:models}

\subsection{Charged Fermions}
\label{sec:model}
We denote the usual $SU(5)$ GUT field content, consisting of 3 generations of $\bf \bar 5$ and  $\bf 10$ each by  
$F^c$ and $A$ respectively, where here and in the following we suppress all flavour indices. 
 
Firstly, we add to this $N$ vectorlike copies $(F,F^c)_{i}$, which are  charged under a vectorlike $U(1)$ symmetry under which $F_i$ has charge $i$ and $F_i^c$ has charge $-i$. The index $i=1\dots N$ will be referred to as a site (in theory space). In addition we introduce the $U(1)$ breaking spurion $\phi$ with charge $-1$. 
We will denote the chiral MSSM GUT fields with an index $i=0$, a notation consistent with their vanishing $U(1)$ charge.
The unique renormalizable superpotential allowed by symmetries is thus of the clockwork type
\be
W_5=\sum_{i=1}^N \phi (F^c_{i-1})^T\C_i F_{i}+(F^c_i)^T(m\A_i+\Sigma \B_i)F_i\,,
\ee
where $m$ is a mass scale, $\Sigma$ is the $SU(5)$ breaking adjoint field, and $\A_i$, $\B_i$ and $\C_i$ are dimensionless couplings that are $3\times 3$ arbitrary complex matrices.\footnote{Throughout this paper we adopt the convention that calligraphic capital letters denote complex $3\times 3 $ matrices.}
The $U(1)$ symmetry is anomaly free (and hence can be gauged) once a conjugate field $\phi^c$ is included, we will assume $\langle \phi\rangle  \gg \langle\phi^c\rangle$ such that we can simply ignore it.\footnote{Including $\phi^c$ allows for the other nearest neighbor interactions, of the kind $\phi^c F_{i+1}^c F_i$. This interaction was included in the model of Ref.~\cite{vonGersdorff:2017iym} and leads to very similar structure for the physical Yukawa couplings.}

Secondly, in a completely analogous manner we introduce copies of the fields transforming in the antisymmetric $\bf 10$ representation, charged under a $U(1)'$ symmetry, with superpotential\footnote{
We use "strength-one" notation for all gauge invariants, i.e.
$F^cF=F^c_aF^a$,
$A^cA=\frac{1}{2}A^c_{ab}A^{ab}$, 
$FAA'=\frac{1}{4}F^a A^{bc}A'^{de}\eps_{abcde}$,
$F^c\Sigma F=F^c_a\Sigma^a_{\ b}F^c$,
$A^c\Sigma A=A^c_{ab}\Sigma^a_{\ d}A^{db}$.
Matrix notation (such as $^T$ and $^\dagger$) is then exclusively reserved for flavor space.
} 
\be 
W_{10}=\sum_{i=1}^{N'}\phi' (A^c_{i})^T\, \C'_i A_{i-1}+(A^c_i)^T(m'\A'_i+\Sigma\,\B'_i)A_i\,,
\ee

Furthermore, we denote the two Higgs doublets as $H^c$ and $H$. The Yukawa superpotential reads
\be
W_Y=H^c(A_0)^T\, \mathcal Y F_0^c + \frac{1}{2} H(A_0)^T \Y'A_0\,,
\ee
with $\Y'^T=\Y'$ and $\M_\nu^T=\M_\nu$.
In terms of MSSM fields, this gives
\be
W_Y=H_d (Q_0)^T \Y D^c_0 + H_d (E^c_0)^T \Y\, L_0 +  H_u (Q_0)^T \Y'\,U_0^c+\dots\,,
\ee
where the ellipsis denotes terms with the Higgs triplet.

The field content of our model is summarized in table \ref{tab:fields}.

\begin{table}[thb]
\centering
\begin{tabular}{c|ccccc|ccccc}
	\toprule
&	\multicolumn{5}{c|}{Matter Fields}&\multicolumn{5}{c}{Higgs Fields}\\
		& $F_i^c$	& $F_i$	& $A_i^c$	& $A_i$	& $N^c$& $\Sigma$ & $H^c$ & $H$	& $\phi$	& $\phi'$\\
		& \footnotesize$0\leq i\leq N$& \footnotesize$1\leq i\leq N$& \footnotesize$1\leq i\leq N'$& \footnotesize$0\leq i\leq N'$&\\
	\midrule
$SU(5)$	&$\bf\overline 5$&$\bf5$	&$\bf\overline {10}$	&$\bf 10$&$\bf 1$ &$\bf 24$	&  	$\bf\overline 5$&$\bf5$	&			$\bf 1$	&$\bf 1$				\\
$U(1)$	& $-i$& $i$&0&0&0&0& 0&0& $-1$&0\\
$U(1)'$	&0&0& $-i$& $i$&0&0&0&0&$0$&$+1$\\
	\bottomrule
	\end{tabular}
\caption{Field content and symmetries of the model. All matter fields and their vectorlike partners, $F_i$, $F_i^c$, $A_i$, $A^c_i$, and $N^c$ carry an additional generation index (not shown).}
\label{tab:fields}
\end{table}

The complete superpotential of the charged fermions is $W=W_5+W_{10}+W_Y$.
Integrating out the clockwork fields with $i\neq 0$ exactly, the superpotential becomes $W=W_Y$, while the K\"ahler potential turns into
\be
K=(F_0^c)^\dagger \Z F^c_0+ (A_0)^\dagger \Z' A_0\,,
\label{eq:kahler}
\ee
with 
\be
\mathcal Z\equiv\sum_{i=0}^N (\mathcal \Q_i\cdots \Q_1)^\dagger (\Q_i\cdots \Q_1)
\,,\qquad 
\Q_i\equiv (m\A_i+v_{24}Y\B_i)^{-1}\phi\,\C_i\,,
\ee
and analogously for $\Z'$. Here $Y$ is the hypercharge (with SM normalization, $Y_Q=1/6$). Notice that the flavor matrices $\Z$ become hypercharge dependent and thus provide a source of GUT breaking. It is this effect that we want to exploit in order to separate the down quark and charged lepton Yukawa couplings.
After canonical normalization, the physical Yukawa couplings read
\be
 \Y_u^*=(\E_Q)^T \Y' \E_{U^c}\,,\qquad  
 \Y_d^*=(\E_Q)^T \Y \E_{D^c}\,,\qquad
 \Y_e^*=(\E_{L})^T\Y^T \E_{E^c}\,,
\ee
where the Hermitian matrices $\E_X$ are defined as 
\be
\E_{D^c}\equiv(\Z_{{2}/{3}})^{-{1}/{2}}\,,\qquad
\E_{L}\equiv(\Z_{-{1}/{2}})^{-{1}/{2}}\,,\ee
and
\be
\E_{Q}\equiv(\Z'_{{1}/{6}})^{-1/2}\,,\qquad
\E_{U^c}\equiv(\Z'_{-{2}/{3}})^{-1/2}\,,\qquad
\E_{E^c}\equiv(\Z'_{1})^{-{1}/{2}}\,.
\ee
Here the subscripts on the $\Z$, $\Z'$ refer to hypercharge. The eigenvalues of $\Z$ ($\E$) are always larger (smaller) than one. 

Assuming no further relation between couplings it is reasonable to expect that the matrices $\A^{(\prime)}$, $\B^{(\prime)}$,  $\C^{(\prime)}$, and $\Y^{(\prime)}$ have $\mathcal O(1)$ complex entries.
As has been shown \cite{vonGersdorff:2017iym,deSouza:2019wji}, the matrices $\E_X$, even though not containing any a priori large or small parameters,
spontaneously develop strongly hierarchical spectra, i.e., their three eigenvalues  satisfy 
\be
\eps_{X^1}\ll\eps_{X^2}\ll\eps_{X^3}\leq 1\,.
\ee 
We will refer to this kind of hierarchies as inter-generational hierarchies, to distinguish them from inter-species hierarchies between the different matter representations (e.g.~between top and tau).
In order to better understand this spontaneous generation of inter-generational hierarchies, let us define  positive parameters
\be
a\equiv \frac{m}{\phi}\,,\qquad b\equiv \frac{v_{24}}{\phi}\,,
\ee
and
\be
a'\equiv \frac{m'}{\phi'}\,,\qquad b'\equiv \frac{v_{24}}{\phi'}\,.
\ee
In the extreme limit $a+b\to \infty$, we have 
\be
\Z\approx1\,,
\label{eq:lim1}
\ee
and there are no inter-generational hierarchies,
while in the opposite limit of $a+b\to 0$, we obtain a product structure
\be
\Z\approx (\mathcal \Q_N\cdots \Q_1)^\dagger (\Q_N\cdots \Q_1)\,,
\label{eq:lim2}
\ee
which spontaneously generates large inter-generational hierarchies between the three eigenvalues. 
This can be understood in terms of a general property of products of random $O(1)$ matrices \cite{vonGersdorff:2017iym}.
The parameters $a$ and $b$ thus interpolate between a hierarchical and non-hierarchical situation.
It is noteworthy that the hierarchies in Eq.~(\ref{eq:lim2}) become independent of  $a,b$ (in the sense that only the eigenvalue's overall size but not  their ratios depend on them).
Interestingly, even in the case $a+b\sim 1$,~\footnote{More precisely $a+|Y| b\sim 1$.}
a strong hierarchy is still present, especially between the third and second generations (see also the discussion in \cite{deSouza:2019wji}).
 
A complementary explanation of this mechanism can be provided by the localization of the zero modes. As has also been shown in ref.~\cite{deSouza:2019wji}  the zero modes of the matter fields spontaneously localize sharply around some random site in theory space, a property first pointed out in \cite{Craig:2017ppp} in similar models.
In this interpretation, the zero modes of the three generations are localized at different sites in theory space, which explains their hierarchical overlap with the site-zero fields.
The parameters $a$ and $b$  set a bias for this localization, the larger $a$ and $b$, the more the zero modes are localized towards site zero.

Moving to a basis in which the $\E_X$ are diagonal, the Yukawa couplings assume the structure $(\Y_u)_{ij}\sim \eps_{Q_i}\eps_{U^c_j}$ etc, familiar from Froggatt-Nielsen \cite{Froggatt:1978nt}, extra-dimensional \cite{Grossman:1999ra,Gherghetta:2000qt,Huber:2000ie}, or strongly coupled \cite{Nelson:2000sn} models, and the hierarchies of masses and mixings follow in a way similar to those (see for instance Ref.~\cite{vonGersdorff:2019gle}).
The CKM angles scale as $\theta_{ij}\sim\eps_{Q^i}/\eps_{Q^j}$ ($i<j$) and similarly for the PMNS anlges with $Q\to L$. The charged fermion masses on the other hand scale as $\eps_{Q^i}\eps_{U^i}$, $\eps_{Q^i}\eps_{D^i}$, and $\eps_{L^i}\eps_{E^i}$, respectively.
Then, the CKM hierarchies roughly determine the $\eps_{Q^i}$ hierarchies. As a general rule, this saturates the hierarchies in the down quark masses, i.e., the required hierarchies in the $\eps_{D^i}$ are rather mild, while the hierarchies in the up quark masses require further suppression from the $\eps_{U^i}$. In the lepton sector, clearly one must avoid a large hierarchy in the $\eps_{L^i}$ in order to keep the PMNS angles large. The charged lepton hierarchies then come mostly from the $\eps_{E^i}$ parameters. From these  rough considerations alone, it is already quite apparent that the structure is rather $SU(5)$ compatible (large hierarchies in the $\bf 10$ sector and mild hierarchies in the $\bf \bar 5$ sector). In section \ref{sec:sims} we will indeed  show that the $SU(5)$ model works remarkably well.
However, it turns out that our ideas are even partially successful in $SO(10)$ unified models, and it is worthwhile then to develop an $SO(10)$ version of the model, which will be done in section \ref{sec:so10}.




\subsection{Neutrinos}

To generate neutrino masses we will employ the seesaw mechanism \cite{Minkowski:1977sc,Yanagida:1979as,Mohapatra:1979ia,Ramond:1979py,GellMann:1980vs,Schechter:1980gr}. We will assume no clockwork fields for the neutrinos, such that we have the simple superpotential
\be
W_\nu=H_u (N^c)^T \Y'' L_0 +\frac{1}{2}(N^c)^T \M N^c+\dots\,,
\ee
where we have already discarded the Higgs triplets. Integrating out the right handed (RH) neutrinos as well as the clockwork leptons gives the Weinberg operator
\be
W_{\nu}=-\frac{1}{2}(H_u L)^T  (\E_L^T\Y''^T\M^{-1}\Y''\E_L)\,(H_uL)\,.
\ee
We parametrize $\M=m_R \, \D$, where $m_R$ is a scale and $\D$ is a further dimensionless order one complex (symmetric) matrix. 

\section{An $SO(10)$ extension.}
\label{sec:so10}

It is possible to extend the previous model to $SO(10)$. The SM fields (including the right handed neutrino) unify into a spinorial $\bf 16$ representation, denoted by $S$.
To break   $SO(10)$ down to the SM one needs more than one irreducible representation. We will choose a $\bf 45$ and a vectorlike ($\bf 126+\overline {126}$).
These are the lowest-dimensional representations that satisfy the following three criteria: $(i)$ breaking of $SO(10)\to$ SM , $(ii)$ possibility to write Yukawa couplings in the clockwork Lagrangian with $\bf 16$, $\bf \overline{16}$ and a GUT breaking Higgs field, $(iii)$ possibility to write Majorana neutrino masses \cite{Ramond:1979py,GellMann:1980vs}.\footnote{Criterion $(i)$ allows also for other low-dimensional representations, for instance $\bf 54$ instead of $\bf 45$ and $\bf 16$ instead of $\bf 126$. However criterion $(ii)$ selects $\bf 45$ over $\bf 54$, and criterion $(iii)$ selects $\bf 126$ over $\bf 16$.}

\subsection{Charged Fermions}

\begin{table}[thb]
\centering
\begin{tabular}{c|cc|ccccc}
	\toprule
&	\multicolumn{2}{c|}{Matter Fields}&\multicolumn{5}{c}{Higgs Fields}\\
&		 $S_i^c$	& $S_i$	& $\Sigma$  & $\Xi^c$ &$\Xi$ & $H$	& $\phi$	\\
&		 \footnotesize$1\leq i\leq N$& \footnotesize$0\leq i\leq N$ &&\\
	\midrule
$SO(10)$	&$\bf\overline {16}$	&$\bf 16$ &$\bf 45$ & $\bf \overline{126}$ & $\bf 126$	&$\bf 10$	&		$\bf 1$				\\
$U(1)$	&$-i$&$i$&0&0& 0&0&$1$\\
	\bottomrule
	\end{tabular}
\caption{Field content and symmetries of the $SO(10)$ model. All matter fields ($S_i$, $S^c_i$) carry an additional generation index (not shown).}
\label{tab:fields2}
\end{table}

We are then lead to consider a single $U(1)$ symmetry (the combination $Q-Q'$ of the symmetries in the $SU(5)$ model which commutes with $SO(10)$). The field content is given in table \ref{tab:fields2}.
The model is defined by 
\be 
W_{16}=\sum_{i=1}^{N}\phi\, (S^c_{i})^T \C_i\, S_{i-1}+(S^c_i)^T(m\A_i+\Sigma\,\B_i)\, S_i\,,
\ee
along with the unique Yukawa coupling
\be
W_Y=\frac{1}{2}H(S_0)^T\Y S_0\,.
\ee

\begin{table}[!htb]
\centering
\begin{tabular}{ccccc}
	\toprule

$\Sigma$	& $\mathbb H_{45}$ & $\mathbb H_{45}\cap \mathbb H_{125}$ & Yukawa relations 	\\
	\midrule
$B-L$		& $SU(3)\times SU(2)_L\times SU(2)_R\times (B-L)$& SM		& $\Y_u\stackrel{*}{=}\Y_d$, $\Y_e=\Y_\nu$\\
$4Y-5(B-L)$	& $SU(5)\times [4Y-5(B-L)]$						& $SU(5)$	& $\Y_e\stackrel{*}{=}(\Y_d)^T$\\
$2Y-(B-L)$	& $SU(4)\times SU(2)_L\times [2Y-(B-L)]$		& SM		& $\Y_d\stackrel{*}{=}\Y_e$, $\Y_u=\Y_\nu$\\
$4Y+(B-L)$	& $SU(5)'\times [4Y+(B-L)]$						& SM		& $\Y_\nu=(\Y_u)^T$\\
else		& SM$\times (B-L)$								& SM		& none\\
	\bottomrule
	\end{tabular}
\caption{Breaking patterns of $\bf45 +126+\overline{126}$. Note that $\mathbb H_{126}=SU(5)$ always. The Yukawa relations marked with $*$ are ruled out.}
\label{tab:so10}
\end{table}

The VEV of the $\bf 45$ is a generic linear combination of hypercharge and $B-L$ generators, which breaks $SO(10)$ down to $\mathbb H_{45}=$ SM$\times (B-L)$. At the same time the VEV of the $\bf 126$ representation breaks $SO(10)$ to $\mathbb H_{126}=SU(5)$, leaving as the unbroken subgroup $\mathbb H_{45}\cap\mathbb H_{126}=\rm SM$. 
In certain particular directions of $\langle \Sigma\rangle$, the group $\mathbb H_{45}$ can be enhanced. 
These directions are summarized in table \ref{tab:so10}.
Since only the {\bf 45} VEV participates in the GUT breaking of the Yukawa couplings, an enhanced $\mathbb H_{45}$ may lead to some relations between Yukawa couplings even when $\mathbb H_{45}\cap \mathbb H_{126}=\rm SM$.
This is the case for all but the last row in table \ref{tab:so10}.\footnote{It is worthwhile to point out that the first row in table \ref{tab:so10} corresponds to the Dimopoulos-Wilczek mechanism \cite{} which is therefore impossible to realized within our model.}

Let us write the VEV of $\Sigma$ as
\be
\langle \Sigma \rangle = v_{45} [\sin\alpha\, Y+\cos\alpha\, (B-L)]\,.
\label{eq:v45}
\ee
The model then has one discrete and 4 continuous non-stochastic parameters: $N$, $a\equiv m/\phi$, $b\equiv v_{45}/\phi$, $\tan \alpha$, and $\tan\beta$.\footnote{A further parameter, the seesaw scale, only affects the neutrino sector}
According to table \ref{tab:so10}, the values $\tan\alpha=0,\ -\frac{4}{5},\ -2$ are ruled out by either $\Y_u=\Y_d$ or $\Y_d=\Y_e$.

\subsection{Neutrinos}

With the symmetry assignments as in table \ref{tab:so10}, there is a unique Majorana mass term for the RH neutrinos
\be
W_{R}=\frac{1}{2}v_{126}N_0^{cT} \D N_0^{c}\,,
\ee
where $\D$ is another complex order one symmetric, dimensionless matrix.
All other Majorana mass terms are forbidden by the nonzero $U(1)$ charges. Considering the non-hierarchical nature of neutrino masses, this is actually very welcome, since this implies that the hierarchical factors $\E_\nu$ drop out of the Weinberg operator. 
We could, for instance, first integrate out the clockwork gauge-singlets, and the RH neutrino fields $N^c_0$ would obtain the usual hierarchical wave function renormalization factor. But the normalization of the RH neutrino's kinetic term  is irrelevant, as they are integrated out in the see-saw mechanism. 

\section{Simulations}
\label{sec:sims}

In this section we would like to find a sets of parameters such that the mechanism leads to a successful generation of the SM flavor structure.
We distinguish two kind of  model parameters. The first kind quantify some underlying physics assumption, such as the scales of symmetry breaking, or the number of vectorlike fields.  
The following parameters are of this type: 
\be
N,\ N',\ a,\ b,\ a',\ b',\ \tan\beta,\ m_R\qquad SU(5) {\rm \ model,}
\ee
\be
N,\ a,\ b,\ \tan\alpha,\ \tan\beta,\ m_R\qquad SO(10) {\rm \ model.}
\ee
We will refer to these parameters as non-stochastic or deterministic.

The remaining parameters are the coupling matrices $\A^{(\prime)}_i$, $\B^{(\prime)}_i$, $\C^{(\prime)}_i$, $\Y^{(\prime)}$ and $\D$. In the absence of additional structure, such as symmetries that would constrain the form of these matrices, it is natural to assume that they have order-one complex entries. 
Choosing them randomly from some suitable prior distribution defines an ensemble of models (with the same physics assumptions). We can then compute the distributions of physical observables (masses and mixings), for each choice of the deterministic parameters.
In order to model the property "order one" for the matrix elements, we will chose flat uniform priors with $|\operatorname{Re}(\A_i)_{kl}| \leq 1$ and $|\operatorname{Im}(\A_i)_{kl}| \leq 1$ etc. Of course, the "posterior distributions" depend to some extent on the choice of priors.\footnote{We have checked though that the posterior that for instance the $\chi^2$ values reported in tables \ref{tab:simu1} and \ref{tab:simu2} are virtually identical if we substitute the uniform priors with Gaussians.}

In order to quantify the success of the mechanism, we proceed as follows.
Let us define the variables 
\be
x_i\equiv \log_{10} O_i\,,
\ee
where the observables $O_i$ run over the nine physical Yukawa couplings, the three CKM mixing angles $\theta_{ij}$, as well as the three quantities $\sin^2\theta_{ij}$ of the PMNS matrix.~\footnote{For the time being we will focus our attention to these 15 observables (all taken at the GUT scale). We will comment on the remaining observables (neutrino mass squared differences and CP violating phases) below.}
It turns out that the logarithms of the observables
 roughly follow a multi-dimensional Gaussian distribution. It is therefore useful to approximate this distribution by a Gaussian with mean and covariance 
taken from the exact (simulated) distribution.
This defines a $\chi^2$ function
\be
\chi(x_i)=(x_i-\bar x_i)(x_j-\bar x_j)C^{-1}_{ij}\,,
\ee  
where the means $\bar x_i$ and covariances $C_{ij}$ depend on the deterministic parameters. This $\chi^2$ function can be used to quantify how the experimental point $x_i^{\rm exp}$ compares to the typical models in the ensemble.
We use as experimental input the values given in table \ref{tab:exp}.~\footnote{This should be seen as a representative case only, as the precise values depend on the supersymmetric threshold corrections. Furthermore, we can ignore the experimental uncertainties which are completely negligible with respect to the width of the theoretical distribution.} Instead of $\chi^2(x_i^{\rm exp})$, an equivalent but  more meaningful quantity to look at is the associated "\pv", the proportion of models that have a larger $\chi^2$  than the experimental point, that is, which are less likely. In our context, a \pv\ $\sim 1$  indicates that the experimental value roughly coincides with the mean of the theoretical distribution, implying that the ensemble typically features a SM-like flavor structure. 

One can then  optimize the deterministic parameters in order to yield larger \pvs, that is, the SM point belongs to the most likely models of the ensemble (it sits near the mean of the distribution).
We should make a disclaimer though to avoid misconceptions. We are not performing a usual $\chi^2$ fit of a model to experimental data. Rather, we are optimizing the deterministic parameters such that the theoretical distributions of models have the SM point as a typical outcome. To quantify this statements we use $\chi_{\rm models}^2(x_{i}^{\rm exp})$ (and the associated \pvs) as a measure.

We will take a look at the $SU(5)$ and $SO(10)$ cases separately.

\subsection{$SU(5)$}

\begin{figure}
\includegraphics[width=\linewidth]{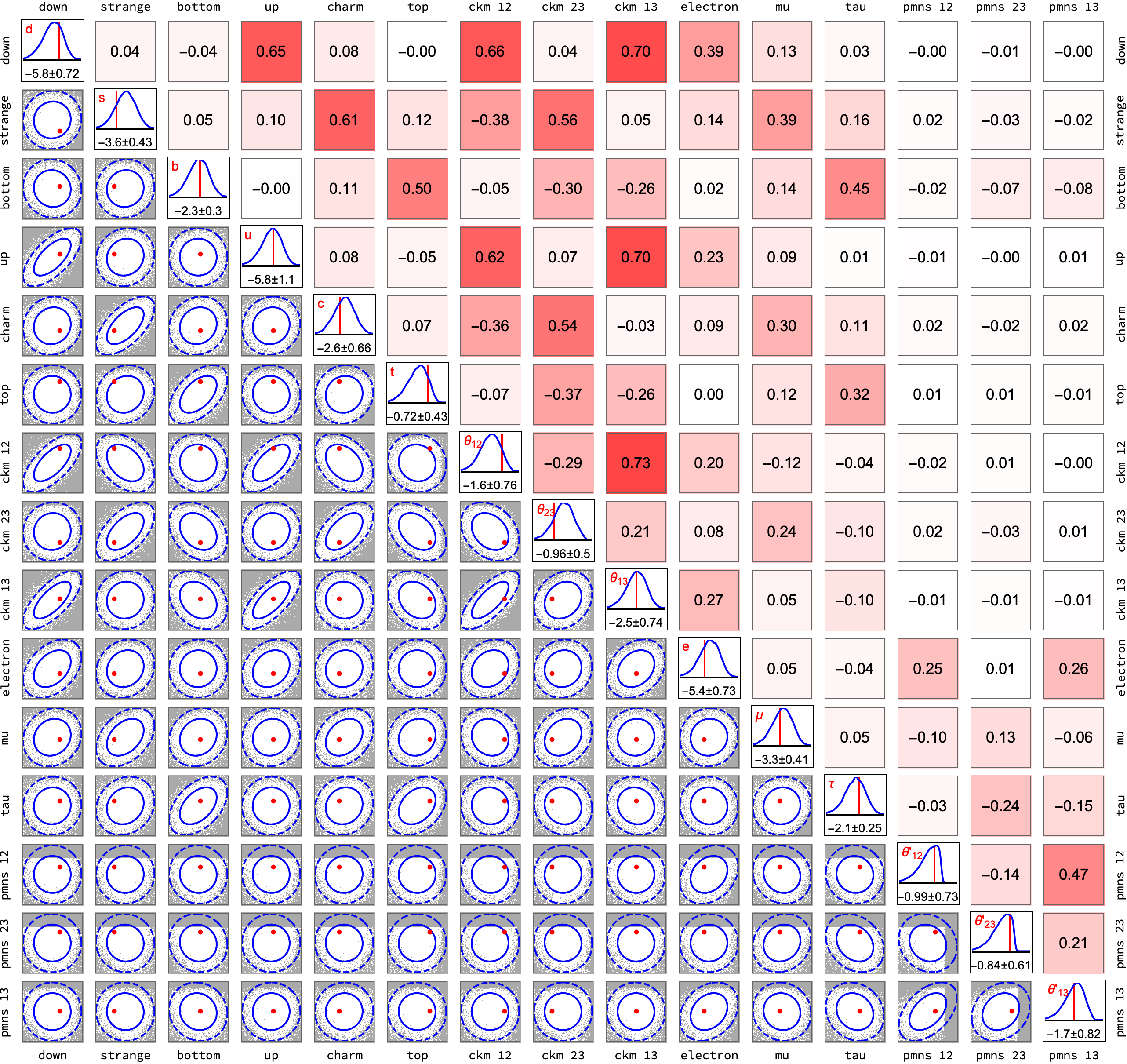}
\caption{The distribution of masses and mixings (parameters $x_i$) in the $SU(5)$ model, for the case $N=1$, $N'=5$ and $\tan\beta=40$. In the lower triangle, we display scatter plots of the two-dimensional marginal distributions, with the solid (dashed) contour representing one (two) sigma, and the red dots the experimental value. In the diagonal, we show the one dimensional marginal distributions over a 3 sigma range with the red lines indicating the experimental value. The numbers on the bottom of each entry are the mean and standard deviation of the theoretical distributions. In the upper triangle, we show the correlation coefficients.}
\label{fig:su5dist}
\end{figure}

We will consider two benchmark values, $\tan\beta=40$ and $\tan\beta=10$. 
For each pair of $N,N'$, we can then optimize the continuous parameters $a$, $b$, $a'$ and $b'$. However, roughly speaking, the optimal values of $a,b$ ($a',b'$) only depend on $N$ ($N'$) and not $N'$ ($N$). Therefore, we can group the parameters as shown in table \ref{tab:simu1} and \ref{tab:simu2}. 
A more refined tuning of the continuous parameters can lead to slightly smaller $\chi^2$ for some values of $(N,N')$, but we don't believe that this adds anything to the general conclusions.
We also give in figure \ref{fig:su5dist} the distributions for the case $N=1$, $N'=5$.
Several features are worth pointing out.
\begin{itemize}
\item
The \pvs\ in tables \ref{tab:simu1} and \ref{tab:simu2} can be quite close to one, especially in the $\tan\beta=40$ case, indicating that our mechanism results in very SM-like masses and mixings. 
\item
As evident from Fig.~\ref{fig:su5dist}, only weak correlations between the lepton and down sectors (for instance between mu and strange masses) persist, due to the GUT breaking effects built into the clockwork Lagrangian.
One easily realizes  large differences such as $m_\mu/m_s\sim 5$.
\item
The breaking of the degeneracy between $\Y_d$ and $\Y_e$ comes mostly from the $\bf \bar 5$ and not from the $\bf 10$ sector (the fit prefers $a<b$ and $a'>b'$). The reason is that the  hierarchies  must mainly come from the $\bf 10$ sector, as explained at the end of section \ref{sec:model}, but
the  hypercharges predict larger hierarchies in $\E_{E^c}$ than in $\E_Q$, which goes in the wrong direction. 
Therefore,  the GUT breaking terms proportional to $b'$ cannot be too large, and 
$N\neq 0$ (even if not strictly needed to generate the hierarchies) is crucial to get $\Y_d\neq\Y_e^T$. 
\footnote{ 
For $N=0$, we could find no choices of parameters with $\chi^2<40$ ($p>10^{-4}$).} 
\item
At larger $N$ ($N'$) a too strong inter-generational hierarchy can be mitigated by larger values of $a,b$ ($a',b')$, see eq.~(\ref{eq:lim1}).
\item
On the other hand for $N'<2$, the inter-generational hierarchies are too small, and this cannot be compensated for by going to very small $a',b'$, which in this regime have a universal effect on all three generations (see eq.~(\ref{eq:lim2}) and the discussion there).
\item
There exist some deviations from the Gaussian approximation for the leptonic observables $\sin^2\theta_{12}$ and $\sin^2\theta_{23}$, due to the fact that their distributions peak near the upper limit $\sin^2\theta_{ij}= 1$. 
However, the true probability density at the experimental point is larger than the the one given by the Gaussian approximation, hence our estimate for the global $\chi^2$ is conservative.
\end{itemize}

%

\newcommand{\vtrip}[3]{$\left\{\begin{array}{c}{\bf #1}\\ #2\\ #3\end{array}\right\}$}
\newcommand{\trip}[3]{$\bigl({\boldsymbol #1},\, #2,\, #3\bigr)$}

\begin{table}[thb]
\centering
\begin{tabular}{cc|cccccc}
\toprule
&			&\multicolumn{3}{c}{\trip{N}{a}{b}}\\
&							
			&\trip{1}{0.15}{0.4}	&\trip{2}{0.45}{0.65}&\trip{3}{0.6}{1.0}\\		
\midrule
\multirow{7}{*}{\trip{N'}{a'}{b'}}
	&\trip{1}{0.02}{0.01}	
			& 	32 (0.007)		&  32 (0.007)		& 	33		(0.004)	\\
	&\trip{2}{0.20}{0.10}	
			&	16.8 (0.47)		&	15.7 (0.41)		& 	16.6	(0.35)	\\
	&\trip{3}{0.35}{0.30}	
			&	9.6 (0.85)		&	10.7 (0.77)		&	11.6	(0.71)	\\
	&\trip{4}{0.45}{0.35}	
			&	 7.6 (0.94)	& 	 8.5 (0.90)	&	9.8		(0.83)	\\	 					
	&\trip{5}{0.60}{0.40}	
			&  	 6.7 (0.97)	&	 8.0	(0.92)	&	9.8		(0.84)	\\
	&\trip{6}{0.65}{0.45}	
			&	 6.5 (0.97)	&	 7.6	(0.94)	&	8.9		(0.88)	\\	
	&\trip{7}{0.75}{0.50}	
			&	 6.9	(0.96)	&	 7.8	(0.93)	&	8.9		(0.88)	\\							
\bottomrule	
\end{tabular}
\caption{Simulation for the $SU(5)$ model for the scenario with $\tan\beta=40$. We display the $\chi^2$ of the physical couplings, corresponding to 15 degrees of freedom. The parenthesis give the \pv, the proportion of models with lower probability density (larger $\chi^2$) than the SM. }
\label{tab:simu1}
\end{table}

\begin{table}[thb]
\centering
\begin{tabular}{cc|cccccc}
\toprule
&			&\multicolumn{3}{c}{\trip{N}{a}{b}}\\
&							
			&\trip{1}{0.05}{0.05}	&\trip{2}{0.2}{0.3}&\trip{3}{0.4}{0.3}\\		
\midrule
\multirow{4}{*}{\trip{N'}{a'}{b'}}
	&\trip{3}{0.35}{0.30}	
			&	10.1 (0.81)	&	12.1 (0.67)	&	14.3	(0.5)	\\
	&\trip{4}{0.50}{0.40}	
			&	 8.3 (0.90)		& 	 10.0 (0.82)	&	12.5	(0.64)	\\	 					
	&\trip{5}{0.60}{0.40}	
			&  	 7.5 (0.94)		&	 9.3	(0.86)	&	11.8	(0.69)	\\
	&\trip{6}{0.65}{0.45}	
			&	 7.3 (0.95)		&	 9.0	(0.88)	&	11.7	(0.70)	\\	
\bottomrule	
\end{tabular}
\caption{Simulation for the $SU(5)$ model for the scenario with $\tan\beta=10$. We display the $\chi^2$ of the physical couplings, corresponding to 15 degrees of freedom. The parenthesis give the \pv, the proportion of models with lower probability density (larger $\chi^2$) than the SM. }
\label{tab:simu2}
\end{table}

Before turning to the $SO(10)$ case, let us comment on the remaining observables, namely the mass-squared  differences of the neutrinos and the CP phases in the CKM and PMNS matrices. 
Inverted neutrino mass ordering would imply almost complete degeneracy between the heavier two neutrinos, which is difficult to realize in our models.
Hence normal mass ordering is strongly preferred. In this case, neutrino masses can always be fit rather well. We have checked that  enlarging the set of observables by  $\Delta m_{21}^2$ and $\Delta m_{32}^2$ adds very little to the global $\chi^2$ (after adjusting the see-saw scale $m_R$), approximately $\sim 0.25$ for $N=1$, and $\sim 0.5$ for $N=3$.
However, since there are now two more degrees of freedom, the  \pvs\ are actually slightly higher than those reported in tables \ref{tab:simu1} and \ref{tab:simu2}.
As for the phases $\delta_{\rm CKM}$ and $\delta_{\rm PMNS}$, their distributions are pretty much flat over the entire range $[0,2\pi)$. Since the Gaussian approximation certainly fails for these variables, it is not very meaningful to include them in the analysis above. On the other hand the flatness of the distribution tells us that the experimental values are neither preferred nor disfavored in our class of models.

\subsection{$SO(10)$}

The $SO(10)$ model depends on  $\tan \alpha$, defined in eq.~(\ref{eq:v45}), which parametrizes the relative direction of $SO(10)$ breaking by the $\bf 45$ representation. 
In order to assess the favoured values of $\tan\alpha$, let us 
 define the quantity 
\be
b_{X}\equiv b\left |Y\sin\alpha + (B-L)\,\cos\alpha\right|\,,
\ee
where $X=Q,L,U^c,D^c,E^c$.

The quantity $a+b_X$ controls the average localization of the zero modes of the field $X$: the larger $a+b_X$, the more it is localized towards site zero, while smaller values repel the zero modes from site zero and reduce the couplings to the Higgs.
Ideally, we would like  $b_L$ to be large, as it reduces the hierarchy in the PMNS angles. One has that $b_L$ is the largest amongst the $b_X$ in the regime $-0.8< \tan\alpha< 0$. At the same time we need to avoid to be too close to the end points of this interval, as they correspond to points where some of the Yukawa couplings matrices exactly unify, see table \ref{tab:so10}. A value that works well in this regime is $\tan\alpha\sim -0.6$, which we will use as our main benchmark.
At this point, one has
\be
b_X=b\times \{0.6,\ 0.46,\ 0.34,\ 0.2,\ 0.05
\}
\ee
for $X=L,D^c,E^c,Q,U^c$ respectively.
Since the $a$ and $b$ parameters affect all fields simultaneously, it is no longer possible (contrary to the $SU(5)$ case) to suppress the down type and charged lepton masses by reducing $a,b$ without for instance also reducing the top mass. We must instead resort to large values of $\tan\beta$. We find that a reasonable value is $\tan\beta\sim 50$.

\begin{table}[!htb]
\centering
\begin{tabular}{ccccccc}
	\toprule
	$\boldsymbol N$				&	\bf 3		&	\bf4		&	\bf5	&	\bf6		&	\bf7	&	\bf8	\\
	\midrule
$a$						&	0.3		&	0.4		&	0.5	&	0.6		&	0.7	&	0.7	\\
$b$						&	1.0		&	1.2		&	1.4	&	1.5		&	1.7	&	1.8	\\
\midrule
$\chi^2$				&	24.1	& 19.2		& 17.2	&	16.2	&	16.0&	16.3\\
\pv					& 	0.06	&	0.21	& 0.31	&	0.37	&	0.38&	0.36\\
	\bottomrule
	\end{tabular}
\caption{Simulation for the $SO(10)$ model, with $\tan\alpha=-0.6$ and $\tan\beta=50$. The $\chi^2$ corresponds to 15 degrees of freedom. The row marked \pv\ gives the proportion of models with lower probability density (larger $\chi^2$) than the standard model.}
\label{tab:simu3}
\end{table}

\begin{figure}
\includegraphics[width=\linewidth]{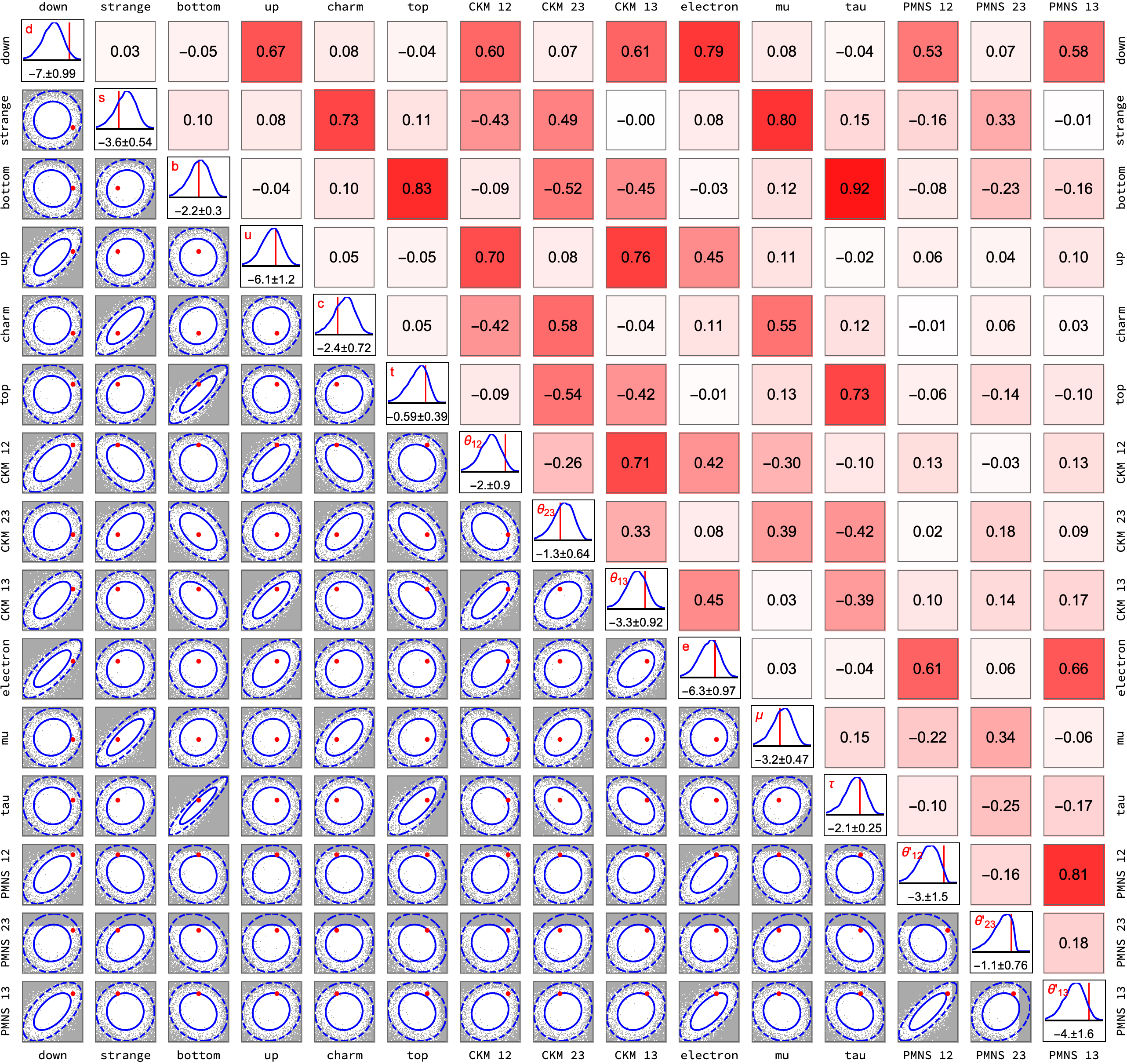}
\caption{The distribution of masses and mixings in the $SO(10)$ model (parameters $x_i$), for the case $N=6$. In the lower triangle, we display scatter plots of the two-dimensional marginal distributions, with the solid (dashed) contour representing one (two) sigma, and the red dots the experimental value. In the diagonal, we show the one dimensional marginal distributions over a 3 sigma range with the red lines indicating the experimental value. The numbers on the bottom of each entry are the mean and standard deviation. In the upper triangle, we show the correlation coefficients.}
\label{fig:so10dist}
\end{figure}

We show in table \ref{tab:simu3} the results of optimizing the remaining parameters $a$ and $b$, and in figure \ref{fig:so10dist} we show the distributions in a representative case.
We can make the following observations.
\begin{itemize}
\item
While certainly less impressive than the $SU(5)$ model, the \pvs\ are nevertheless surprisingly good, considering the necessity of rather large $SO(10)$ breaking effects. 
For  $N\geq 5$, the experimental point $x_i^{\rm \exp}$ is less than one sigma away from the mean of the distribution.
\item
As in the case of $SU(5)$, at large values of $N$, potentially too large inter-generational hierarchies  are partially erased by increasing  $a$ or $b$, which explains why the $\chi^2$ values do not increase again at larger $N$. However, they also do not improve any further for $N>7$.
\item
Neutrinos with normal mass ordering can again be fit easily in our model. The typical increase in $\chi^2$ is of the order of $\sim 0.3$, which for two additional degrees of freedom slightly {\em increases} the \pvs. 
\end{itemize}

\section{Phenomenology}
\label{sec:pheno}

There are two principle impacts on low energy observables that could be used to constrain this class of models.

Firstly, as any supersymmetric GUT model, exchange of the triplet Higgs can mediate proton decay via dimension-five operators.
In minimal $SU(5)$ GUTs, precision gauge coupling unification requires a triplet Higgs mass that allows for proton decay at a rate incompatible with data \cite{Nath:2006ut}. In our model, there are additional vectorlike matter fields with associated threshold corrections. Notice that  only mass ratios enter in the generation of the Yukawa couplings, and the overall mass scale of these new particles is a free parameter. 
 For each model in our scan, one can in principle adjust this mass scale and the triplet mass to achieve precise gauge coupling unification, and calculate the associated proton lifetime. 
  Such an analysis is however not independent of the solution to the doublet triplet splitting problem, and the two issues should be dealt with together. For instance, the missing partner mechanism \cite{Masiero:1982fe,Grinstein:1982um,Hisano:1994fn} typically require extended Higgs sectors.
 Even though we have presented our mechanism for a minimal Higgs sector, we expect it to work similarly well for extended Higgs sectors, as long as we can write some Yukawa interactions of the CW matter fields with the GUT-breaking Higgs fields. On the other hand it would also be interesting to try and take advantage of the CW mechanism itself  to split the doublet and triplet masses. 
 A fully realistic model in this regard, including doublet-triplet splitting and an analysis of proton decay,  is left to future work.
 
 Secondly, we should comment about low energy flavor violating  signatures of these models. The  wave function renormalization factors,  see e.g.~eq.~(\ref{eq:kahler}), will also strongly reduce flavor violation in the soft masses \cite{Nomura:2007ap,Dudas:2010yh}. The effect is virtually identical to strongly coupled \cite{Nelson:2000sn} or extra-dimensional  \cite{Choi:2003fk,Nomura:2008pt} models of supersymmetric flavor, though quite different from models with horizontal symmetries, which are more constrained than the former  \cite{Dudas:2010yh}. One of the most constraining  observables is the decay of $\mu\to e\gamma$. If the charged lepton hierarchy is mostly coming from the $\E_e$ (as in our $SU(5)$ scenario), one has  \cite{Dudas:2010yh}
 \be
\left( \frac{A_0}{100\ \rm GeV}\right)\left(\frac{400\  \rm GeV}{\tilde m_\ell}\right)^4<0.4\,,
 \ee
 where $A_0$ is the  trilinear soft term and $\tilde m_\ell$ the slepton mass (similar bounds have been derived in ref.~\cite{Nelson:2000sn}). 
 When $\E_\ell$ is somewhat hierarchical (as in our $SO(10)$ model), the bounds become  weaker.
 In the quark sector, the strongest bounds come from the neutron EDM which require 
squark masses $\tilde m_q\gtrsim 1$ TeV  \cite{Dudas:2010yh}. Since the sfermion masses are dominated by the gaugino loops, squark and slepton masses are related as $\tilde m_q\sim5 \tilde m_\ell$, and the leptonic bounds are more constraining.

\section{Conclusions}

\label{sec:conclusions}

We have presented a renormalizable GUT model of flavor which accounts very well for the observed hierarchies of masses and mixings in the charged fermion sector.
The model features one and two spontaneously broken $U(1)$ symmetries in the case of $SU(5)$ and $SO(10)$ respectively. Contrary to Froggatt-Nielsen type models, the MSSM chiral matter fields are uncharged under this symmetry. 
We have taken advantage of the GUT breaking terms present in the most general renormalizable CW Lagrangian in order to lift the degeneracy of the down and lepton Yukawa couplings.
Inter-generational hierarchies result spontaneously from products of $\mathcal O(1)$ matrices, while inter-species hierarchies can either arise from a CW-like suppression or from large $\tan\beta$.

In $SU(5)$, for a GUT breaking scale slightly smaller than the $U(1)$ breaking scales we obtain distributions of models that 
feature the SM point amongst the $\sim 5\%$  most likely models, i.e., very close to the mean value of the distribution.
This requires roughly about one vectorlike copy of the $\bf \bar 5$ matter fields, and $\gtrsim 5$ copies of the $\bf 10$.
For the best $SO(10)$ case, the SM fits slightly worse in the distributions, belonging only to about the $60\%$ most likely models, approximately one sigma away from the mean of the distribution. A good fit requires about $N\gtrsim 5$ vectorlike copies of the entire MSSM matter content.

The fact that the probability density is unsuppressed at the SM point can also be interpreted in terms of fine-tuning. Accidental cancellations occur only very rarely at random. We could exclude such points from our distributions, but this would at most affect the far tails of the distributions, implying that there are many non-fine tuned models which reproduce the experimental data of table~\ref{tab:exp} precisely.

\bibliographystyle{JHEP}

\bibliography{paper}
\bibliographystyle{hieeetr}

 
\end{document}